\documentclass{article}

\usepackage{amssymb}
\usepackage[pdftex]{graphicx}

\begin{document}
\title{Ordered States in Heisenberg Pyrochlore Antiferromagnets with Dipole-Dipole Interactions}
\author{Terufumi Yokota\\
{\it College of Engineering, Nihon University}\\
{\it Koriyama, Fukushima, 963-8642, Japan}
}



\maketitle

\begin{abstract}
Classical Heisenberg model on a pyrochlore lattice with both a nearest-neighbor antiferromagnetic interaction and long-range dipole-dipole interactions is investigated by numerically solving the Landau-Lifshitz (LL) equation. The ground state of the model without the dipole-dipole interactions is known to be a classical spin liquid and the system does not order at all temperature because of strong frustration. On the other hand, the ordered state of the model with weak dipolar interactions is known to be the so-called Palmer-Chalker (PC) state, in which a pair of spins and another pair of spins are antiparallel and the former is perpendicular to the latter for each tetrahedron of the lattice. Here ordered states of the model are obtained for various values of the relative strength of the two kinds of the interactions. Besides the spin liquid state and the PC-like state, states that might be considered as glassy or viscous spin liquid, states of period 8, states of period 4, and states of period of 4 on one sublattice and 8 on other three sublattices are realized. Among the states with period 4 or (and) 8, multi-{\bf K} structures are observed.    
\end{abstract}

\newpage
\section{Introduction}
The Heisenberg model on a pyrochlore lattice with an antiferromagnetic nearest-neighbor exchange interaction is frustrated strongly. The lattice is a three-dimensional network of corner-sharing tetrahedra. Each tetrahedron is constructed from four triangles that are the origin of the frustration. As a result of the strong frustration, the system does not order at all temperature and is called classical spin liquid\cite{mc}.

Antiferromagnetic Heisenberg models on a pyrochlore lattice with different nearest-neighbor interactions or further-neighbor interactions have also been investigated. The model with two different nearest-neighbor exchange interactions called ``breathing pyrochlore" has been investigated\cite{bs}. In addition to the spin liquid state, a state similar to the type I antiferromagnetic state\cite{gz} appeared in the nearest-neighbor antiferromagnetic model on the FCC lattice can be realized in the case with both ferromagnetic and antiferromagnetic interactions. The Heisenberg model with both antiferromagnetic nearest-neighbor interaction and ferromagnetic next-nearest-neighbor interaction has been investigated\cite{onk}. Multiple-{\bf K} states appear in this model. The Heisenberg models with further-neighbor interactions have also been investigated resulting in multiple-{\bf K} ordered states at low temperatures\cite{zgz}.  

For rare earth pyrochlores like $\rm{Gd_2 Sn_2 O_7}$ and  $\rm{Gd_2 Ti_2 O_7}$, dipolar interactions are important in addition to the antiferromagnetic nearest-neighbor exchange interaction\cite{mg}. Ordered state in $\rm{Gd_2 Sn_2 O_7}$ was suggested to be the Palmer-Chalker (PC) state in which a pair of spins and another pair of spins for each tetrahedron are antiparallel and parallel to the opposite edge of the tetrahedron\cite{pc}. PC ground state has been numerically confirmed for the long-range dipolar interactions at the level of $10 - 20 \%$ of the exchange interaction\cite{mg2}. On the other hand, ordered state in $\rm{Gd_2 Ti_2 O_7}$ was suggested to be different from the PC state but it is a multi-{\bf K} structure\cite{setal}.

Considering various possible ground states in the Heisenberg model with both the antiferromagnetic nearest-neighbor exchange interaction and the dipolar interactions on the pyrochlore lattice, it would be an interesting problem to study the ground states for the model. In this paper, ordered states in the model are investigated by numerically solving the Landau-Lifshitz (LL) equation for various values of the two kinds of interactions. Besides the spin liquid state for the pure antiferromagnetic nearest-neighbor exchange interaction and the PC(-like) states for a range of relative strength of the two kinds of the interactions, various states including states with multi-{\bf K} structures are observed.

In Sec. 2, the antiferromagnetic Heisenberg model with dipole-dipole interactions on the pyrochlore lattice is introduced. The LL equation for the model is given. A continuous space approximation for the dipole-dipole interaction part of the equation is used to make numerical calculations effectively. Ground states are obtained by numerically solving the LL equation for various values of the two kinds of  interactions in Sec. 3. In Sec. 4, discussion and conclusion are given.

\section{Model and Equation}
Heisenberg antiferromagnet on a pyrochlore lattice with dipole-dipole interactions is described by the Hamiltonian
\begin{equation}
\mathcal{H}=\sum_{<i, j>}J_{ij}{\bf S}_i \cdot{\bf S}_j-\omega\sum_{<i, j>}\frac{3({\bf S}_i \cdot{\bf R}_{ij})({\bf S}_j \cdot{\bf R}_{ij})-{\bf S}_i \cdot{\bf S}_j R_{ij}^2}{R_{ij}^5}\equiv \mathcal{H}_{ex}+\mathcal{H}_{dip}
\end{equation}
where $i$ represents a lattice site of the pyrochlore lattice with the periodic boundary condition. The summation is taken over pairs of the pyrochlore lattice sites and $J_{ij}$ takes a non-negative value $J$ only for nearest-neighbor pairs. A cubic unit cell of the pyrochlore lattice is depicted in Fig. 1 where the size of the unit cell is $4a\times 4a\times 4a$.
\begin{figure}
  \begin{center}
   \includegraphics[width=140mm]{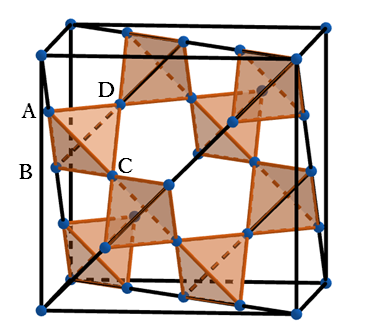}
  \end{center}
\vspace{-0.5cm}
\caption{(Color online) Unit cell of the pyrochlore lattice.}
\end{figure}
In this paper, $a$ is set to $1$. In this figure, four lattice points of a tetrahedron are labeled with A, B, C, and D. All the pyrochlore lattice points can be obtained by a translation along the rectangular coordinate axes and named sublattice A, B, C, and D in this paper. ${\bf S}_i $ is a classical three-dimensional vector spin with unit length. The first term of the Hamiltonian represents the nearest-neighbor antiferromagnetic interaction. The second term of the Hamiltonian represents long-range dipole-dipole interactions where ${\bf R}_{ij}$ is the vector from a site $i$ to a site $j$ and $\rm{R}_{ij} =\vert{\bf R}_{ij}\vert$.

The LL equation is
\begin{equation}
\frac{d{\bf S}_i (t)}{dt}=-{\bf S}_i (t)\times\left({\bf B}_i (t)+\gamma{\bf S}_i (t)\times{\bf B}_i (t)\right)
\end{equation}
where ${\bf B}_i (t)=-\frac{\delta\mathcal{H}}{\delta{\bf S}_i (t)}+{\bf \xi }_i (t)$ and $\gamma$ is the damping constant. ${\bf \xi }_i$ is Gaussian thermal noise, which satisfies
\begin{eqnarray}
&&<{\bf \xi }_i (t)>=0\nonumber\\
&&<\xi^a_i (t)\xi^b_j (t')>=\delta_{ab}\delta_{ij}\delta (t-t')\frac{2\gamma}{1+\gamma^2}T
\end{eqnarray}
where $<\cdots >$ is the average, $T$ is the temperature, and the index $a$ represents the vector component \cite{gl}. $\delta_{ab}$ and $\delta_{ij}$ are the Kronecker deltas and $\delta (t-t')$ is the Dirac delta function.

Fourier transform of the LL equation will be used to properly include the long-range dipole-dipole interactions in the system with periodic boundary condition. A continuous space approximation is used to obtain an explicit form of the dipole-dipole interaction part in the equation.

The exchange part of the equation is
\begin{equation}
{\bf B}_{i, ex} (t)=-\frac{\delta\mathcal{H}_{ex}}{\delta{\bf S}_i (t)}=-\sum_{j (n. n. {\rm of} i)}J{\bf S}_j
\end{equation}
where the summation is taken over the nearest-neighbor sites of $i$. The discrete Fourier transform of the exchange part for the lattice size of $L\times L\times L$ is 
\begin{equation}
{\bf B}_{ex} ({\bf K})=-2J{\bf S}({\bf K})(\cos K_x \cos K_y+\cos K_x \cos K_z+\cos K_y \cos K_z )
\end{equation} 
where ${\bf S}({\bf K})$ is the Fourier transform of ${\bf S}_i$ and $K_A =\frac{2\pi m_A}{L}$. $m_A$ is an integer that satisfies $-\frac{L}{2}<m_A \le \frac{L}{2}$.

The dipole-dipole term for the equations is
\begin{equation}
{\bf B}_{i, dip}=-\frac{\delta\mathcal{H}_{dip}}{\delta{\bf S}_i }.
\end{equation}
The Fourier transform is obtained using the continuous space approximation. The three components of ${\bf B}_{i, dip}$ in the continuous space approximation are given by
\begin{equation}
\label{Bx}
B^x_{i, dip}=\omega\int d{\bf R}'[S^x({\bf R}')G_x({\bf R}-{\bf R}')+3S^y({\bf R}')G_{xy}({\bf R}-{\bf R}')+3S^z({\bf R}')G_{xz}({\bf R}-{\bf R}')],\\
\end{equation} 
\begin{equation}
\label{By}
B^y_{i, dip}=\omega\int d{\bf R}'[S^y({\bf R}')G_y({\bf R}-{\bf R}')+3S^x({\bf R}')G_{xy}({\bf R}-{\bf R}')+3S^z({\bf R}')G_{yz}({\bf R}-{\bf R}')],\\
\end{equation} 
and
\begin{equation}
\label{Bx}
B^z_{i, dip}=\omega\int d{\bf R}'[S^z({\bf R}')G_z({\bf R}-{\bf R}')+3S^x({\bf R}')G_{xz}({\bf R}-{\bf R}')+3S^y({\bf R}')G_{yz}({\bf R}-{\bf R}')]\\
\end{equation} 
where $G_x ({\bf R})=\frac{3x^2-R^2}{R^5}$, $G_y ({\bf R})=\frac{3y^2-R^2}{R^5}$, $G_z ({\bf R})=\frac{3z^2-R^2}{R^5}$, $G_{xy} ({\bf R})=\frac{xy}{R^5}$, $G_{xz} ({\bf R})=\frac{xz}{R^5}$, and $G_{yz} ({\bf R})=\frac{yz}{R^5}$. ${\bf R}=(x, y, z)$ and $R=\vert{\bf R}\vert$.  Discrete Fourier transforms with the form of the continuous space approximation are used to include the long-range dipole-dipole interactions. Explicit forms will be given shortly. The Fourier transforms on the pyrochlore lattice are given by
\begin{eqnarray}
B^x_{dip}({\bf K})&=&\omega\sum_{m'_x , m'_y, m'_z}\epsilon ({\bf K}-{\bf K}')\left[S^x ({\bf K}')G_x ({\bf K}')+3S^y ({\bf K}')G_{xy}({\bf K}')\right.\nonumber\\
&&\hspace{1cm}
\left.+3S^z ({\bf K}')G_{xz}({\bf K}')\right],
\end{eqnarray} 
\begin{eqnarray}
B^y_{dip}({\bf K})&=&\omega\sum_{m'_x , m'_y, m'_z}\epsilon ({\bf K}-{\bf K}')\left[S^y ({\bf K}')G_y ({\bf K}')+3S^x ({\bf K}')G_{xy}({\bf K}')\right.\nonumber\\
&&\hspace{1cm}
\left.+3S^z ({\bf K}')G_{yz}({\bf K}')\right],
\end{eqnarray} 
and
\begin{eqnarray}
B^z_{dip}({\bf K})&=&\omega\sum_{m'_x , m'_y, m'_z}\epsilon ({\bf K}-{\bf K}')\left[S^z ({\bf K}')G_z ({\bf K}')+3S^x ({\bf K}')G_{xz}({\bf K}')\right.\nonumber\\
&&\hspace{1cm}
\left.+3S^y ({\bf K}')G_{yz}({\bf K}')\right],
\end{eqnarray}
where $G({\bf K})$'s are the Fourier transforms of  $G({\bf R})$'s and $\epsilon ({\bf K})$ is
\begin{eqnarray}
\epsilon ({\bf K})&=&\frac{1}{8}\left[2{\bf\delta}_{{\bf K}, {\bf 0}}+2{\bf\delta}_{{\bf K}, {\bf\Pi}}+{\bf\delta}_{{\bf K}, (\frac{\pi}{2}, \frac{\pi}{2}, \frac{\pi}{2})}+{\bf\delta}_{{\bf K}, (-\frac{\pi}{2}, -\frac{\pi}{2}, -\frac{\pi}{2})}+{\bf\delta}_{{\bf K}, (-\frac{\pi}{2}, \frac{\pi}{2}, \frac{\pi}{2})}+{\bf\delta}_{{\bf K}, (\frac{\pi}{2}, -\frac{\pi}{2}, -\frac{\pi}{2})}\right.\nonumber\\
&&\left.+{\bf\delta}_{{\bf K}, (\frac{\pi}{2}, \frac{\pi}{2}, -\frac{\pi}{2})}+{\bf\delta}_{{\bf K}, (-\frac{\pi}{2}, -\frac{\pi}{2}, \frac{\pi}{2})}-{\bf\delta}_{{\bf K}, (\frac{\pi}{2}, -\frac{\pi}{2}, \frac{\pi}{2})}-{\bf\delta}_{{\bf K}, (-\frac{\pi}{2}, \frac{\pi}{2}, -\frac{\pi}{2})}\right].
\end{eqnarray}
${\bf 0}=(0, 0, 0)$ and ${\bf\Pi}=(\pi, \pi, \pi)$. ${\bf\delta}$ is the product of three Kronecker deltas, ${\bf\delta}_{{\bf K}, {\bf 0}}=\delta_{k_x , 0}\delta_{k_y , 0}\delta_{k_z , 0}$, and so on. $G({\bf K})$'s are obtained explicitly performing the integral in the polar coordinate system and given by
\begin{equation}
G_x({\bf K})=4\pi \left[1-3\left(\frac{K_x}{K}\right)^2 \right]\frac{\sin (Kd)-Kd\cos (Kd)}{(Kd)^3},
\end{equation} 
\begin{equation}
G_y({\bf K})=4\pi \left[1-3\left(\frac{K_y}{K}\right)^2 \right]\frac{\sin (Kd)-Kd\cos (Kd)}{(Kd)^3},
\end{equation} 
\begin{equation}
G_z({\bf K})=4\pi \left[1-3\left(\frac{K_z}{K}\right)^2 \right]\frac{\sin (Kd)-Kd\cos (Kd)}{(Kd)^3},
\end{equation} 
\begin{equation}
G_{xy}({\bf K})=-4\pi \frac{K_x K_y}{K^2}\cdot\frac{\sin (Kd)-Kd\cos (Kd)}{(Kd)^3},
\end{equation} 
\begin{equation}
G_{xz}({\bf K})=-4\pi \frac{K_x K_z}{K^2}\cdot\frac{\sin (Kd)-Kd\cos (Kd)}{(Kd)^3},
\end{equation} 
and
\begin{equation}
G_{yz}({\bf K})=-4\pi \frac{K_y K_z}{K^2}\cdot\frac{\sin (Kd)-Kd\cos (Kd)}{(Kd)^3}
\end{equation} 
where $K=\vert {\bf K}\vert$ and $d$ is the cutoff length that is the lower limit for the range of the dipole-dipole interactions. In this paper, $d$ is set to $1$. These continuous space approximations for $G({\bf K})$'s are used as discrete Fourier transforms in Eqs. (10) - (12) for  $K_A =\frac{2\pi m_A}{L}$ with integer $-\frac{L}{2}<m_A \le \frac{L}{2}$.

\section{Various Ordered States}
The LL equation is numerically solved to obtain ground states in this section. Implicit Gauss-Seidel method with fractional step\cite{wge} is used to improve numerical stability. The system size is $64\times 64\times 64$ with the periodic boundary condition that approximates the infinite system. Discrete Fourier transforms with the forms of the continuous space approximation for the dipole-dipole interaction part in Eqs. (14) - (19) are used. In Sec. 4, discussion on the approximation will be given. The damping constant is set as $\gamma =1.0$. The initial directions of the spins are randomly chosen, which corresponds to a paramagnetic state at high temperature. The temperature is set to $5.0$ at the time $t=0.0$ not so as to form a pattern at the initial temperature. The temperature is lowered at a speed of $dT/dt =-10^{-4}$ to $T=0.0$ at $t=50,000$. Spin structure ${\bf S}({\bf R})$ is obtained. The structure may be classified by using Fourier transformed spin variables ${\bf S}({\bf K})$. $\vert S^a ({\bf K})\vert$ has peaks at some wave vectors. The peak values are from several thousands to tens of thousands depending on the relative strength of the exchange and dipolar interactions. In some cases with lower peak values, ordered states might be glassy although definite conclusions cannot be made in the present study. 

\subsection{Spin liquid}
Spin liquid state is realized in the system with only antiferromagnetic exchange interaction without dipolar interactions\cite{mc}. For the numerical solutions of $J=2.0$ and $\omega =0.0$, the total spin of four sites in each tetrahedron is ${\bf 0}$ for the pyrochlore lattice\cite{mc}, which is verified numerically. The sublattice spin correlation functions $\vert g^\alpha (R_{ij}\ne 0)\vert\equiv\overline{\vert{\bf S}_i \cdot{\bf S}_j \vert}$ are obtained, where $R_{ij}=\vert{\bf R}_i -{\bf R}_j \vert$ and $\alpha$ represents sublattice label A, B, C, or D explained in Fig. 1. The overline is the average over ${\vert{\bf S}_i \cdot{\bf S}_j \vert}$ with definite distance $R_{ij}$ between ${\bf S}_i$ and ${\bf S}_j$. If the direction between ${\bf S}_i$ and ${\bf S}_j$ is random and without correlation,  $\vert g^\alpha (R_{ij}\ne 0)\vert =1/2$ and the standard deviation $\sigma =1/(2\sqrt{3})\simeq 0.289$. For the calculated system, the averaged sublattice correlation functions over $R_{ij}$ are $0.504\sim 0.506$ depending on $\alpha$ and $\sigma\simeq 0.29$ although some deviations can be seen with $R_{ij}$. Almost random directions between the two spin directions are consistent with the spin liquid state.

Spin autocorrelation function defined by $A(t)\equiv\frac{1}{N}\sum_i {\bf S}_i (0)\cdot{\bf S}_i (t)$ is depicted in Fig. 2, where $N$ is the total number of spins.
\begin{figure}
  \begin{center}
   \includegraphics[width=140mm]{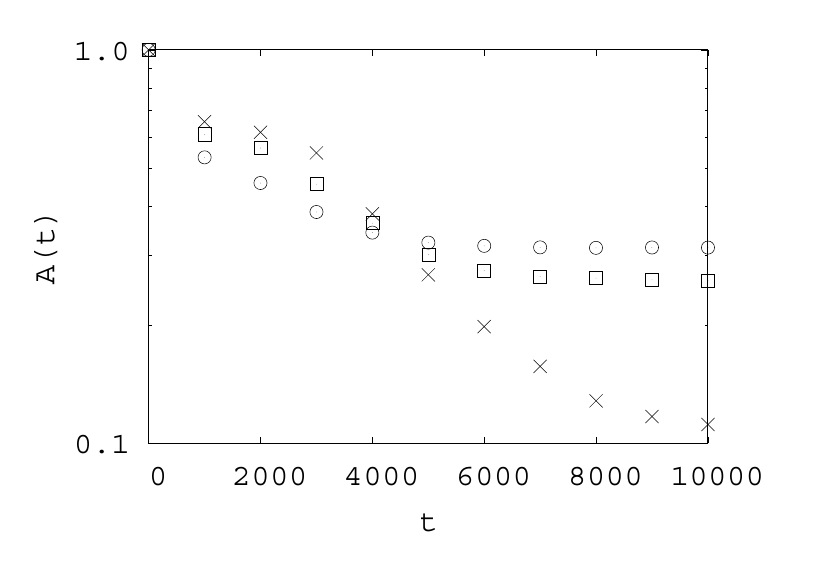}
  \end{center}
\vspace{1.5cm}
\caption{Spin autocorrelation functions. Crosses, squares, and circles are the data for $\omega =0.0, 0.01$, and $0.02$ , respectively with $J=2.0$.}
\end{figure}
In this figure, $t=0$ and $t=10,000$ correspond to $T=1.0$ and $T=0.0$, respectively. The data for $\omega =0.01$ and $0.02$ with $J=2.0$ are added to compare the result for $\omega =0.0$ with those for small non-zero dipolar interactions. The data for $\omega =0.0, 0.01$, and $0.02$ are shown by crosses, squares, and circles, respectively. The results of $\omega =0.0$ seem to behave differently from those of $\omega =0.01$ and $0.02$ for $T<0.5$, which corresponds to $t>5,000$.  $A(t)$ for $\omega =0.0$ at $T=0.0$ is much smaller than those for $\omega =0.01$ and $0.02$. This is consistent with the spin liquid nature of the system with $\omega =0.0$. $A(t)$ for $\omega =0.01$ and $0.02$ does not change for $T\le 0.2$ up to two and three decimal places, respectively. The dynamics of the system with nonzero $\omega$ is much slower than that with $\omega =0.0$.

\subsection{Glassy spin state or viscous spin liquid} 
As noted in the previous subsection, ordered states for $\omega =0.01$ and $0.02$ with $J=2.0$ might be frozen, which might be seen in the behavior of the autocorrelation function $A(t)$ in Fig. 2. The sublattice spin correlation functions averaged over $\vert{\bf S}_i \cdot{\bf S}_j \vert$ with definite distance $R_{ij}$ between ${\bf S}_i$ and  ${\bf S}_j$ and their standard deviations are $\vert g^\alpha (R_{ij}\ne 0)\vert\pm\sigma\simeq 0.545\sim 0.549\pm 0.293\sim 0.295$ and $0.562\sim 0.570\pm 0.295$ for $\omega =0.01$ and $0.02$, respectively. The variations in the average and the standard deviation $\sigma$ come from sublattice $\alpha$. These values are a little bit larger than that for the random case and there is little correlation.

The distributions of  angles between two spin directions, which belong to the same tetrahedron are depicted in Fig. 3.
\begin{figure}
 \begin{minipage}{0.49\hsize}
  \begin{center}
   \includegraphics[width=60mm]{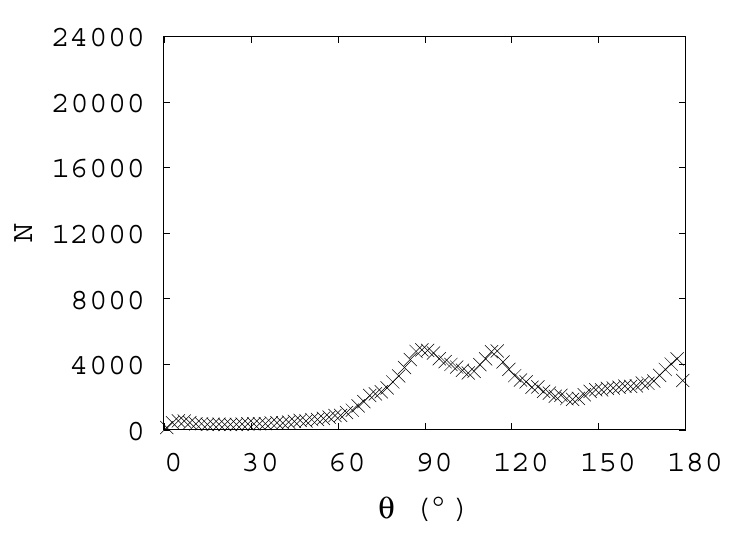}
  \end{center}
 \end{minipage}
 \begin{minipage}{0.49\hsize}
  \begin{center}
   \includegraphics[width=60mm]{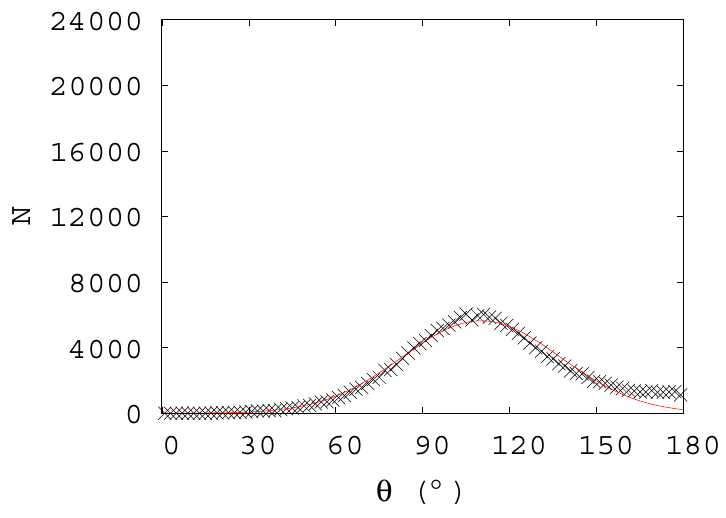}
  \end{center}
 \end{minipage}
\vspace{1.0cm}
\caption{(Color online) Distributions of angles between two spin directions for $\omega =0.0$ (left) and $\omega =0.01$ (right) with $J=2.0$.}
\end{figure}
In the figure, the distributions for  $\omega =0.0$ (left) and $\omega =0.01$ (right) with $J=2.0$ are shown. The distribution for $\omega =0.01$ is different in shape from that for $\omega =0.0$, which suggests different kinds of spin directions between the two systems. The Gaussian fitting curve is added for $\omega =0.01$. The distribution for $\omega =0.02$ is similar to that for $\omega =0.01$. There is a broad peak near $110^\circ$ and the standard deviation is about $24^\circ$ for $\omega = 0.01$. The angle $\arccos (-1/3)\simeq 109.5^\circ$ corresponds to all-in and all-out magnetic order which is realized in a spin ice model with the nearest-neighbor antiferromagnetic exchange interaction and the dipolar interactions\cite{mg3}. The all-in and all-out spin structure is suggested also in the Heisenberg model with Dzyaloshinsky-Moriya interaction\cite{eetal}.

Although the spin orientations for $\omega =0.01$ and $0.02$ have a character of all-in and all-out magnetic order on average, the deviations are rather large. The orientations of spins belonging to a tetrahedron have almost no correlations with those belonging to other  tetrahedra and they are not related to the orientation of the tetrahedron, which is manifested by the broad peak near $110^\circ$ in Fig. 3. Although the state for $\omega =0.01$ and $0.02$ might be frozen one, possibility of spin liquid state also for $\omega =0.01$ and $0.02$ cannot be ruled out. After reaching zero temperature, calculations have been continued to extra time interval $\Delta t=10,000$ at $T=0.0$. Spin autocorrelation functions at $T=0.0$ are depicted in Fig.4.
\begin{figure}
  \begin{center}
   \includegraphics[width=140mm]{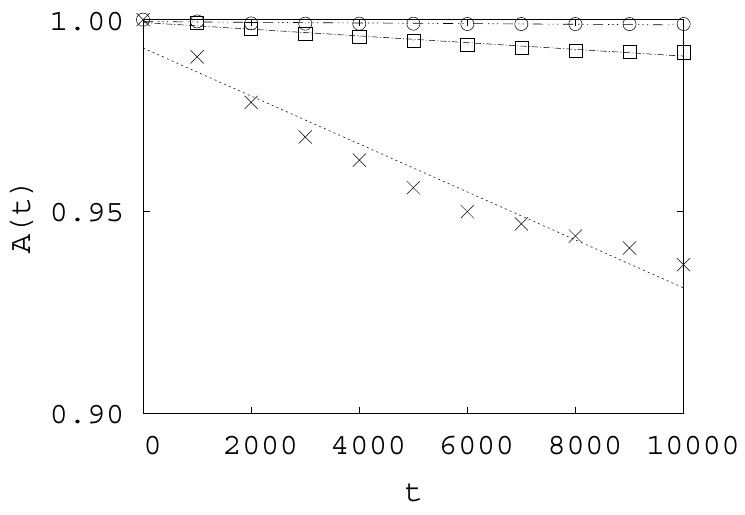}
  \end{center}
\vspace{1.5cm}
\caption{Spin autocorrelation functions at $T=0.0$. Crosses, squares, and circles are the data for $\omega =0.0, 0.01$, and $0.02$ , respectively with $J=2.0$.}
\end{figure}
In this figure, $t=0$ corresponds to $t=10^4$ in Fig.2. Logarithmic scale is used for $A(t)$. The data for $\omega =0.0$, $0.01$, and $0.02$ are shown by crosses, squares, and circles, respectively. In this time period, the autocorrelation function is approximated as $A(t)\simeq ae^{bt}$. The data for each $\omega$ are approximated with a line in the figure. The fitting parameters $a$ and $b$ are $a\simeq 0.99$ and $b\simeq -6.4\times 10^{-6}$ for $\omega =0.0$, $a\simeq 1.00$ and $b\simeq -9.0\times 10^{-7}$ for $\omega =0.01$, and $a\simeq 1.00$ and $b\simeq -8.8\times 10^{-8}$ for $\omega =0.02$, respectively. This suggests approximate time scales of the autocorrelation decay is about $10^6$ for $\omega =0.01$ and about $10^7$ for $\omega =0.02$. The time scales are much larger than the time $10^4$ of the present simulation at $T=0.0$. Definite conclusion whether the state is frozen or liquid with slow dynamics cannot be reached and further studies are needed.  

\subsection{(Quasi) Palmer-Chalker (PC) states}
In this subsection, systems with several $\omega /J$'s are investigated. For the PC state, a pair of spins and another pair of spins in each tetrahedron are antiparallel and parallel to the opposite edge of the tetrahedron\cite{pc}.

For $\omega /J =0.125$ with $\omega =0.5$ and $J=4.0$, the PC state is realized. In Fig. 5, spin orientations in a tetrahedron are shown.
\begin{figure}
  \begin{center}
   \includegraphics[width=140mm]{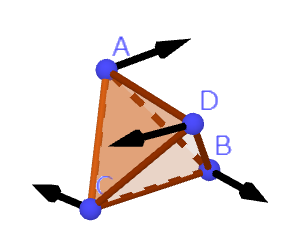}
  \end{center}
\vspace{-0.5cm}
\caption{(Color online) Spin orientations in a tetrahedron for $\omega =0.5$ and $J=4.0$.}
\end{figure}
The sublattice spin correlation function $g^\alpha (R_{ij})=1.0$ and perfect order is realized. The distribution of the angles between two spin directions $N(\theta )$ has sharp peaks around $\theta =90^\circ$ and $175^\circ$. The latter represents almost antiparallel spin directions. In Ref. 4, PC ground state is shown to be realized at the level of $0.1\lesssim\omega /J\lesssim 0.2$, which is consistent with the present result. Spin orientations in the same region of $xy$ and $xz$ planes for two successive $z$ layers and $y$ layers, respectively from left to right are depicted in Fig. 6.
\begin{figure}
 \begin{minipage}{0.05\hsize}
  \begin{center}
   \includegraphics[width=7mm]{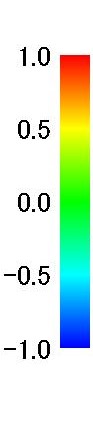}
  \end{center}
 \end{minipage}
 \begin{minipage}{0.22\hsize}
  \begin{center}
   \includegraphics[width=28mm]{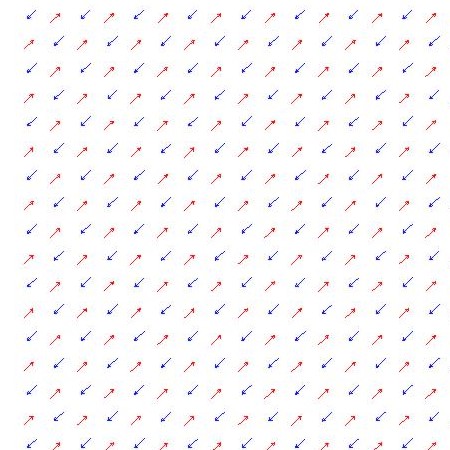}
  \end{center}
 \end{minipage}
 \begin{minipage}{0.22\hsize}
  \begin{center}
   \includegraphics[width=28mm]{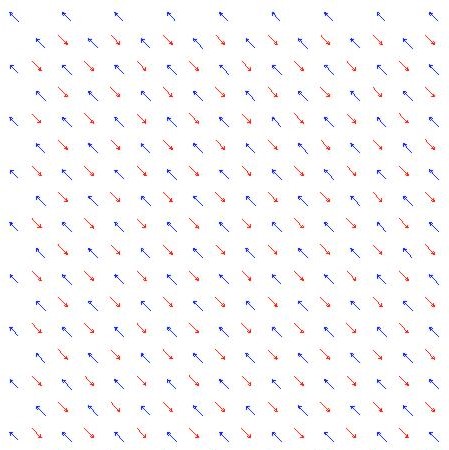}
  \end{center}
 \end{minipage}
 \begin{minipage}{0.22\hsize}
  \begin{center}
   \includegraphics[width=28mm]{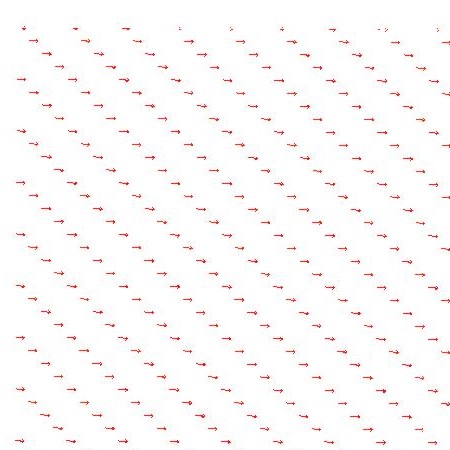}
  \end{center}
 \end{minipage}
 \begin{minipage}{0.22\hsize}
  \begin{center}
   \includegraphics[width=28mm]{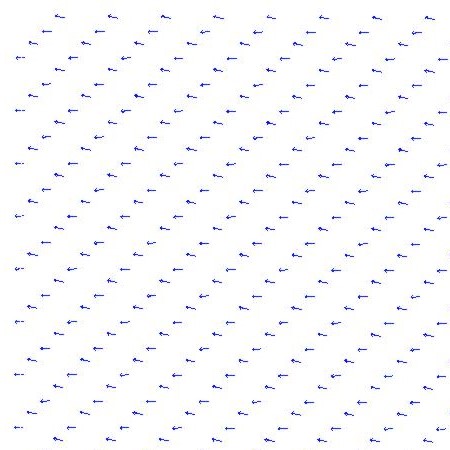}
  \end{center}
 \end{minipage}\vspace{1.5cm}
\vspace{-1.5cm}
\caption{(Color online) Spin orientations in the same regions of $xy$ and $xz$ planes, respectively for two successive $z$ layers and $y$ layers from left to right  for $\omega =0.5$ and $J=4.0$. The color legend shows the value of $S^x$.}
\end{figure}
The color legend shows the value of $S^x$.

For smaller $\omega /J$'s, that is $\omega /J=0.1, 0.05$, and $0.025$ with $J=2.0$, quasi-PC states appear. $N(\theta )$'s for these systems are depicted in Fig. 7.
\begin{figure}
 \begin{minipage}{0.32\hsize}
  \begin{center}
   \includegraphics[width=40mm]{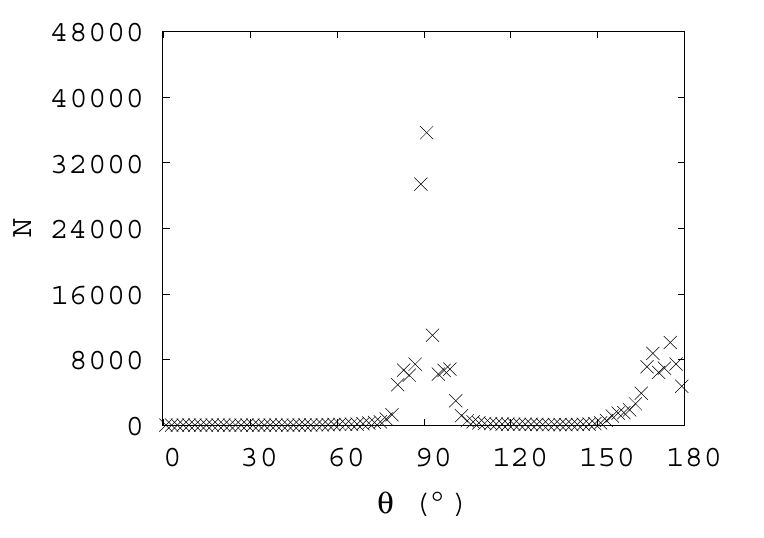}
  \end{center}
 \end{minipage}
 \begin{minipage}{0.32\hsize}
  \begin{center}
   \includegraphics[width=40mm]{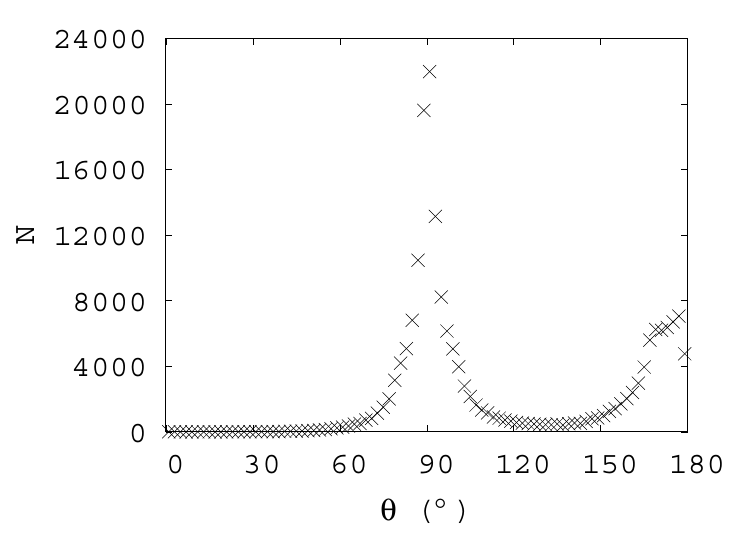}
  \end{center}
 \end{minipage}
 \begin{minipage}{0.32\hsize}
  \begin{center}
   \includegraphics[width=40mm]{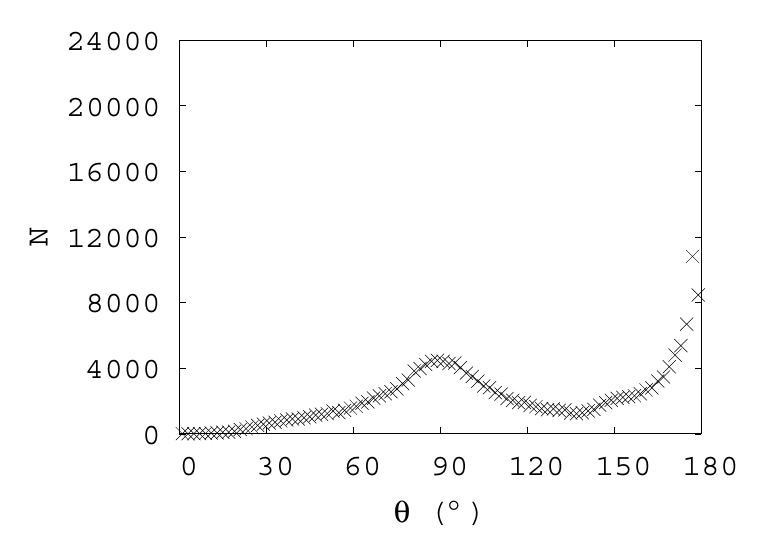}
  \end{center}
 \end{minipage}
\vspace{0.5cm}
\caption{Distributions of angles between two spin directions for $\omega =0.2, 0.1$ and $0.05$ (from left to right) with $J=2.0$.}
\end{figure}
For $\omega /J =0.025$, the peaks around $\theta =90^\circ$ and $177^\circ$ are broad. Two peaks for $\omega /J =0.05$ are not so broad and those for $\omega /J  =0.1$ around $\theta =90^\circ$ and $175^\circ$ are narrower. Spin orientations in the same regions of $xy$ planes for successive $z$ layers for $\omega /J =0.1$ are depicted in Fig. 8.
\begin{figure}
 \begin{minipage}{0.49\hsize}
  \begin{center}
   \includegraphics[width=57mm]{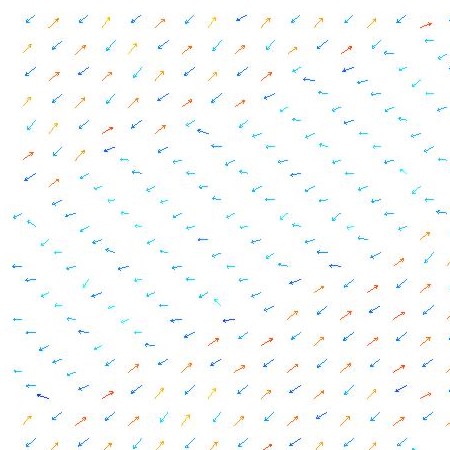}
  \end{center}
 \end{minipage}
 \begin{minipage}{0.49\hsize}
  \begin{center}
   \includegraphics[width=57mm]{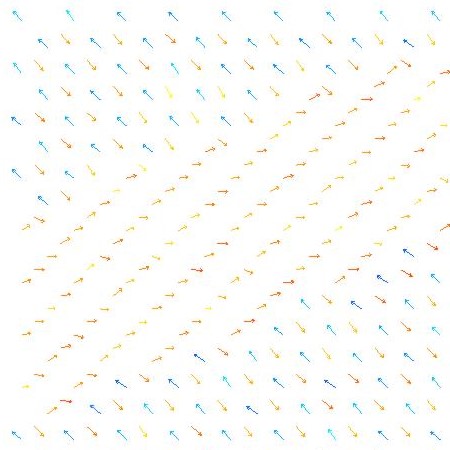}
  \end{center}
 \end{minipage}
\vspace{0.5cm}
\caption{(Color online) Spin orientations in the same regions of xy planes for successive $z$ layers for $\omega /J =0.1$. The color scale is identical to that in Fig. 5.}
\end{figure}
The color scale is identical to that in Fig. 6. Compared with Fig. 6, there are stripe domains with the alternate spin orientations similar to those in $z$ and $y$ layers in Fig. 6. Fluctuations in the spin orientation within a domain are larger than those for $\omega /J =0.125$ depicted in Fig. 6. Stripe domain patterns appear also in ferromagnetic Heisenberg models with dipole-dipole interactions although the spin structures in these systems are simpler\cite{y1}\cite{y2}. Spin orientations in the same regions of $xy$ planes for successive  $z$ layers for $\omega /J =0.05$ (left) and $\omega /J =0.025$ (right) are depicted in Fig. 9.
\begin{figure}
 \begin{minipage}{0.24\hsize}
  \begin{center}
   \includegraphics[width=28mm]{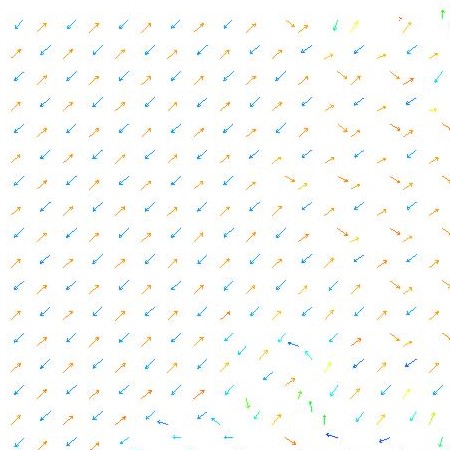}
  \end{center}
 \end{minipage}
 \begin{minipage}{0.24\hsize}
  \begin{center}
   \includegraphics[width=28mm]{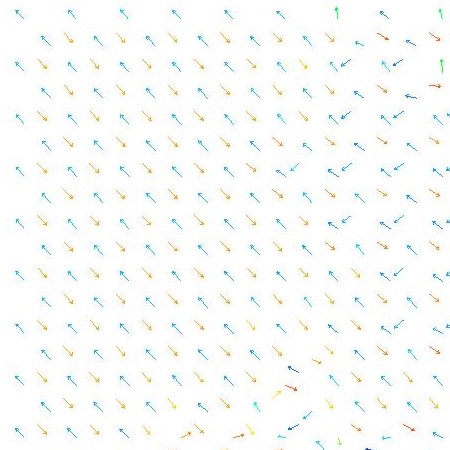}
  \end{center}
 \end{minipage}
 \begin{minipage}{0.24\hsize}
  \begin{center}
   \includegraphics[width=28mm]{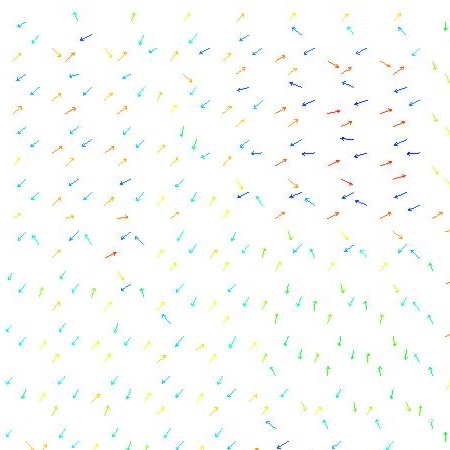}
  \end{center}
 \end{minipage}
 \begin{minipage}{0.24\hsize}
  \begin{center}
   \includegraphics[width=28mm]{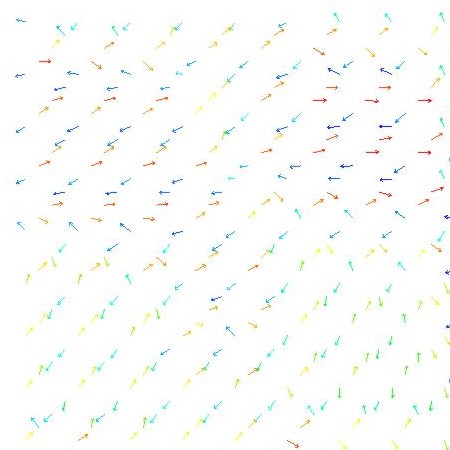}
  \end{center}
 \end{minipage}\vspace{1.5cm}
\vspace{-1.5cm}
\caption{(Color online) Spin orientations in the same regions of $xy$ planes for two successive $z$ layers for $\omega /J=0.05$ (left) and $\omega /J=0.025$ (right). The color scale is identical to that in Fig. 6.}
\end{figure}
The color scale is identical to that in Fig. 6. There are irregular shaped domains with the nature of the PC state for $\omega /J =0.05$. As can be seen in Figs. 7 and 9, the spin orientations show character of quasi-PC state.

Also for larger $\omega /J$, that is $\omega /J =0.25$ with $J=2.0$, a quasi-PC state appears. $N(\theta )$ and spin orientations of successive $z$ layers for $\omega /J =0.25$ are depicted in Fig. 10.
\begin{figure}
 \begin{minipage}{0.32\hsize}
  \begin{center}
   \includegraphics[width=40mm]{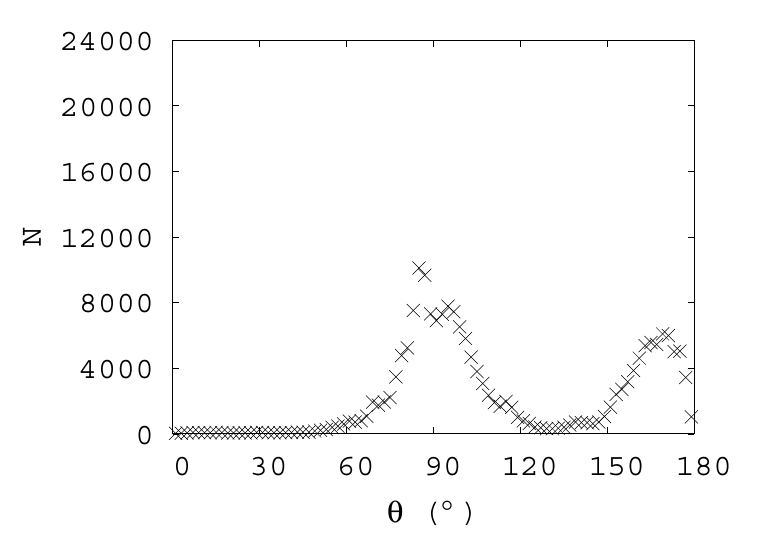}
  \end{center}
 \end{minipage}
 \begin{minipage}{0.32\hsize}
  \begin{center}
   \includegraphics[width=38mm]{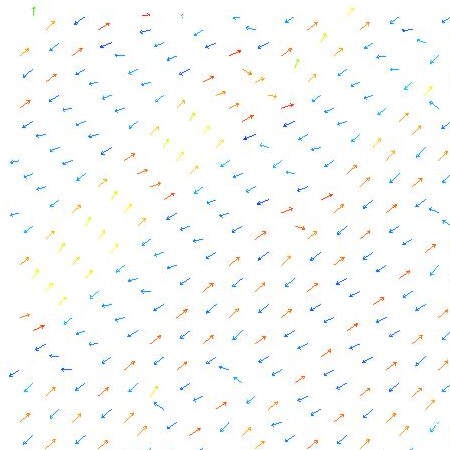}
  \end{center}
 \end{minipage}
 \begin{minipage}{0.32\hsize}
  \begin{center}
   \includegraphics[width=38mm]{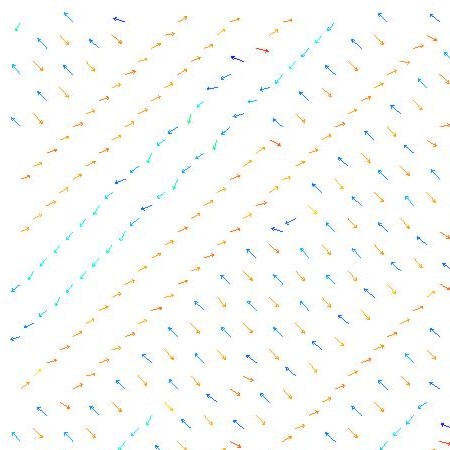}
  \end{center}
 \end{minipage}
\vspace{0.5cm}
\caption{(Color online) Distributions of angles between two spin directions (left) and spin orientations in the same regions of the $xy$ planes for two successive $z$ layers for $\omega /J=0.25$ with $J=2.0$. The color scale is identical to that in Fig. 6.}
\end{figure}
The color scale is identical to that in Fig. 6. There are two broad peaks around $\theta =85^\circ$ and $170^\circ$ which are a little smaller than the values for the PC ground state. There are stripes with irregular narrow width.  

\subsection{Multiple-{\bf K} states with period 8}
For larger values of $\omega /J$ than the previous subsection, spin structures with approximate period 8 are observed. In Fig. 11, the same regions in $xy$ planes of successive $z$ layers for $\omega /J =1.0$ with $J=0.5$ are depicted to show the spin orientations.
\begin{figure}
 \begin{minipage}{0.24\hsize}
  \begin{center}
   \includegraphics[width=28mm]{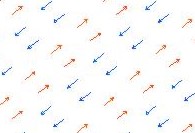}
  \end{center}
 \end{minipage}
 \begin{minipage}{0.24\hsize}
  \begin{center}
   \includegraphics[width=28mm]{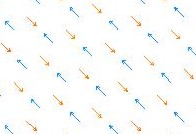}
  \end{center}
 \end{minipage}
 \begin{minipage}{0.24\hsize}
  \begin{center}
   \includegraphics[width=28mm]{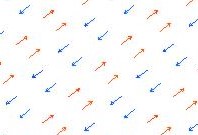}
  \end{center}
 \end{minipage}
 \begin{minipage}{0.24\hsize}
  \begin{center}
   \includegraphics[width=28mm]{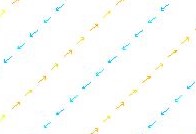}
  \end{center}
 \end{minipage}\vspace{1.5cm}
\vspace{-1.5cm}
\caption{(Color online) Spin orientations in the same region in $xy$ planes of successive $z$ layers for $\omega /J =1.0$ with $J=0.5$. The color scale is identical to that in Fig. 6.}
\end{figure}
The color scale is identical to that in Fig. 6. The spin structure can be classified by using Fourier transformed spin variables ${\bf S}({\bf K})$. The peaks of $\vert S^a ({\bf K})\vert$ are largest for $a=x$ and they are at ${\bf K}\simeq\pi (1/4, -1/4, \pm 1/4)$, $\pi (1/4, 3/4, \pm 1/4)$, $\pi (3/4, 1/4, \pm 3/4)$, and $\pi (3/4, -3/4, \pm 1/4)$ within the deviation of $\pi\cdot\delta =\frac{2\pi}{64}=\frac{2\pi}{L}$ for each component of the wave vector, which seems to show octuple-{\bf K} structure. $\pi\cdot\delta$ is the minimum deviation in the $L=64$ system. Approximate period 8 spin structure can be seen in the figure. Similar spin structures are observed for $\omega /J =0.5$ and $2.5$ with $\omega =0.5$.

Although spin structures are different, cases of $\omega /J =5.0$ and $10.0$ with $\omega =0.5$ show approximate period 8. They are shown in Figs. 12 and 13, respectively.
\begin{figure}
 \begin{minipage}{0.24\hsize}
  \begin{center}
   \includegraphics[width=28mm]{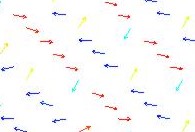}
  \end{center}
 \end{minipage}
 \begin{minipage}{0.24\hsize}
  \begin{center}
   \includegraphics[width=28mm]{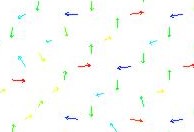}
  \end{center}
 \end{minipage}
 \begin{minipage}{0.24\hsize}
  \begin{center}
   \includegraphics[width=28mm]{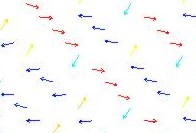}
  \end{center}
 \end{minipage}
 \begin{minipage}{0.24\hsize}
  \begin{center}
   \includegraphics[width=28mm]{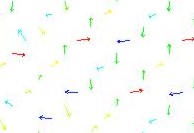}
  \end{center}
 \end{minipage}\vspace{1.5cm}
\vspace{-1.5cm}
\caption{(Color online) Spin orientations in the same region in $xy$ planes of successive $z$ layers for $\omega /J =5.0$ with $\omega=0.5$. The color scale is identical to that in Fig. 6.}
\end{figure}
\begin{figure}
 \begin{minipage}{0.24\hsize}
  \begin{center}
   \includegraphics[width=28mm]{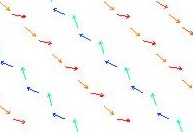}
  \end{center}
 \end{minipage}
 \begin{minipage}{0.24\hsize}
  \begin{center}
   \includegraphics[width=28mm]{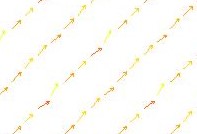}
  \end{center}
 \end{minipage}
 \begin{minipage}{0.24\hsize}
  \begin{center}
   \includegraphics[width=28mm]{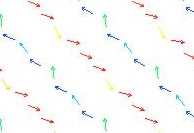}
  \end{center}
 \end{minipage}
 \begin{minipage}{0.24\hsize}
  \begin{center}
   \includegraphics[width=28mm]{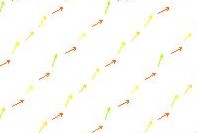}
  \end{center}
 \end{minipage}\vspace{1.5cm}
\vspace{-1.5cm}
\caption{(Color online) Spin orientations in the same region in $xy$ planes of successive $z$ layers for $\omega /J =10.0$ with $\omega=0.5$. The color scale is identical to that in Fig. 6.}
\end{figure}
The same regions in the $xy$ planes  of successive $z$ layers are depicted in both figures. The color scale is identical to that in Fig. 6. The deviations from the period 8 can be seen in both figures especially for $\omega /J =5.0$. The wave vectors which are the peaks of  $\vert S^a ({\bf K})\vert$ indicate approximate period 8 spin structures. The largest peak value of  $\vert S^a ({\bf K})\vert$ for $\omega /J =5.0$ is $1.14\times 10^4$, which is somewhat smaller than $1.67\times 10^4$ for $\omega /J =2.5$. The multiplicity for the ordered state is not clear for  both $\omega /J =5.0$ and $10.0$.

\subsection{Coexistence of ferromagnetic order with period 4 and 8 state}
Ordered states in pure dipolar systems depend on crystal structures. For the pyrochlore lattice, the ordered state for $\omega =0.5$ and $J=0.0$ is rather complicated as shown below. The peaks of  $\vert S^a ({\bf K})\vert$ are largest for $a=x$ and they are at ${\bf K}=\pi (0, 0, \delta)$, ${\bf K}\simeq\pi (1/4, 1/4, -1/4)$, $\pi (1/4, 1/4, 3/4)$, $\pi (3/4, 3/4, 1/4)$, and $\pi (3/4, 3/4, -3/4)$, and ${\bf K}\simeq\pi (1/2, -1/2, 1/2)$ within the deviation of $\pi\cdot\delta$. ${\bf K}=\pi (0, 0, \delta)$ indicates a ferromagnetic order with two domains which are related with inversion of the magnetization. In Fig. 14, the number distributions of $S^x ({\bf R})$ on sublattices A to D are depicted.
\begin{figure}
 \begin{minipage}{0.24\hsize}
  \begin{center}
   \includegraphics[width=28mm]{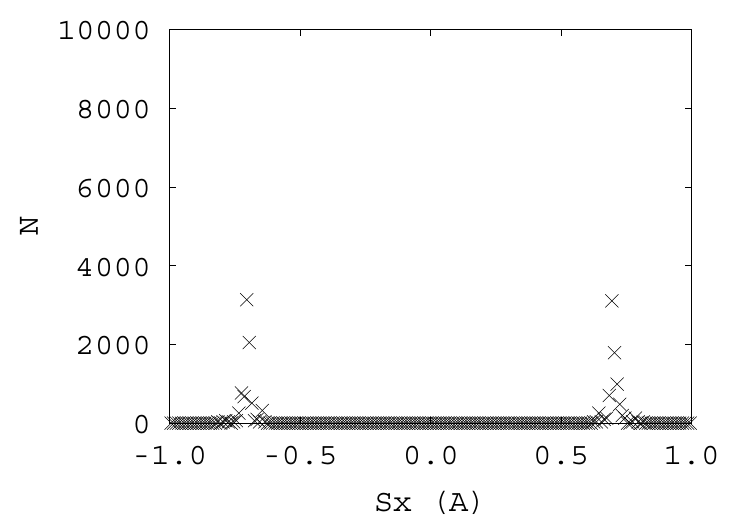}
  \end{center}
 \end{minipage}
 \begin{minipage}{0.24\hsize}
  \begin{center}
   \includegraphics[width=28mm]{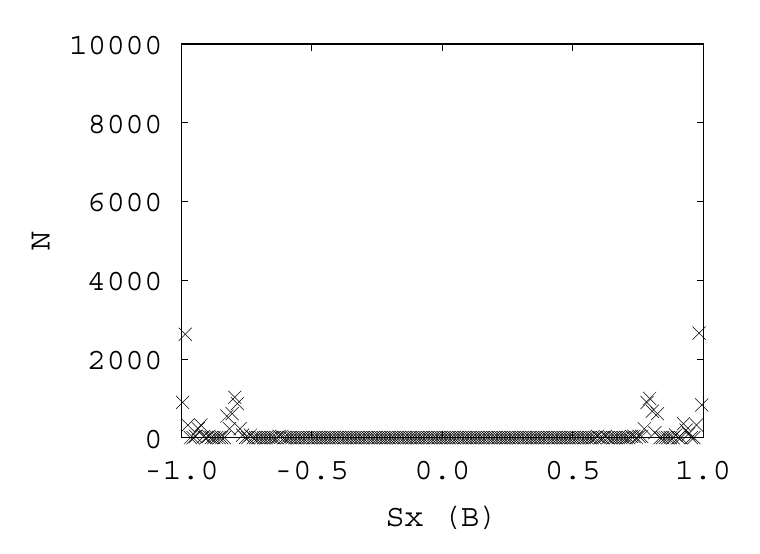}
  \end{center}
 \end{minipage}
 \begin{minipage}{0.24\hsize}
  \begin{center}
   \includegraphics[width=28mm]{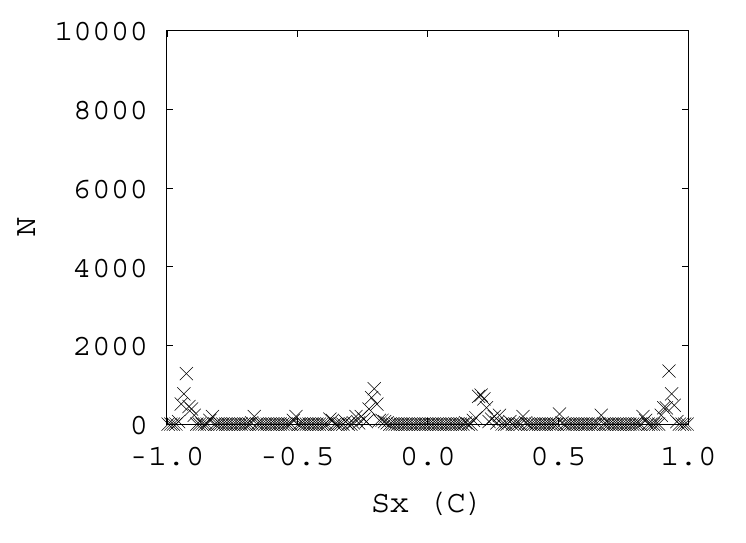}
  \end{center}
 \end{minipage}
 \begin{minipage}{0.24\hsize}
  \begin{center}
   \includegraphics[width=28mm]{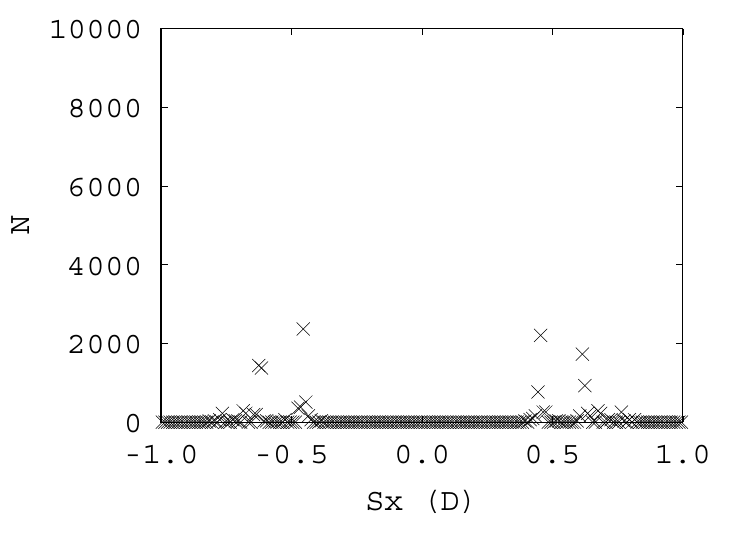}
  \end{center}
 \end{minipage}\vspace{1.5cm}
\vspace{-1.0cm}
\caption{Number distributions of $S^x ({\bf R})$ for $\omega =0.5$ and $J=0.0$ at sublattices A to D from left to right.}
\end{figure}
Except for the approximate symmetry of the distribution $N(S^x )\simeq N(-S^x )$, there is one peak for sublattice A whereas there are two peaks for sublattice B, C, and D. This corresponds to the spin structure of period 4 for the sublattice A and that of period 8 for the sublattices B, C, and D. The periodicity also appears in the peaks of $\vert S^a ({\bf K})\vert$ mentioned above. Peak at $\pi (1/2, -1/2, 1/2)$ represents period 4 and those at $\pi (1/4, 1/4, -1/4)$ and $\pi (1/4, 1/4, 3/4)$, $\pi (3/4, 3/4, 1/4)$, and $\pi (3/4, 3/4, -3/4)$ represent period 8, respectively. In Fig. 15,  the same region in $xy$ planes of successive $z$ layers for $\omega =0.5$ and $J=0.0$ is depicted to show the spin orientations.
\begin{figure}
 \begin{minipage}{0.24\hsize}
  \begin{center}
   \includegraphics[width=28mm]{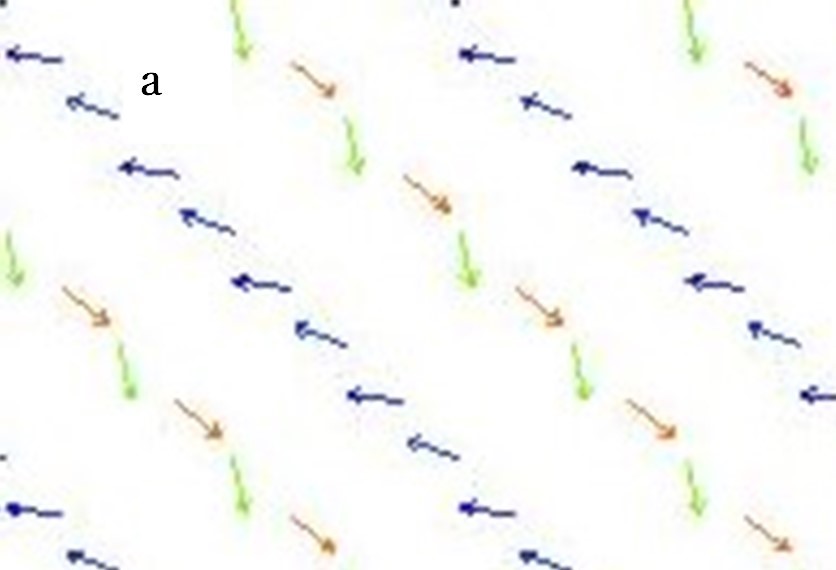}
  \end{center}
 \end{minipage}
 \begin{minipage}{0.24\hsize}
  \begin{center}
   \includegraphics[width=28mm]{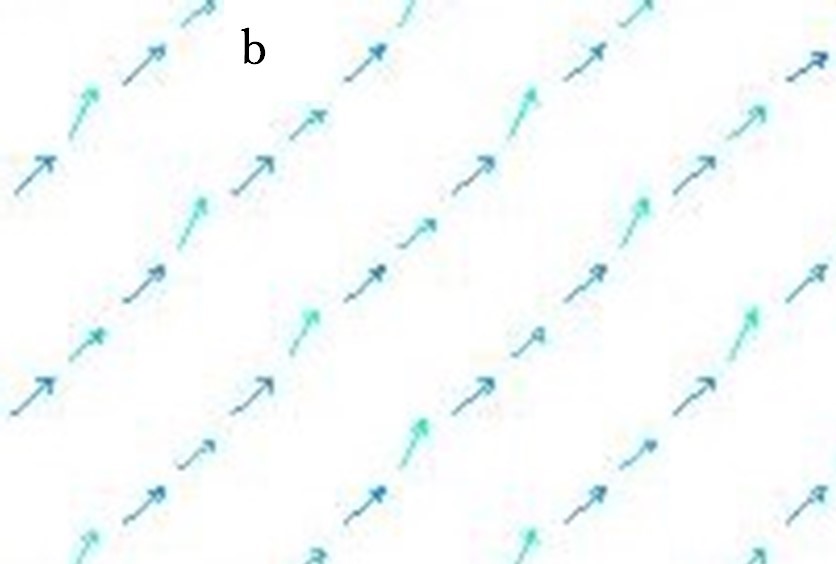}
  \end{center}
 \end{minipage}
 \begin{minipage}{0.24\hsize}
  \begin{center}
   \includegraphics[width=28mm]{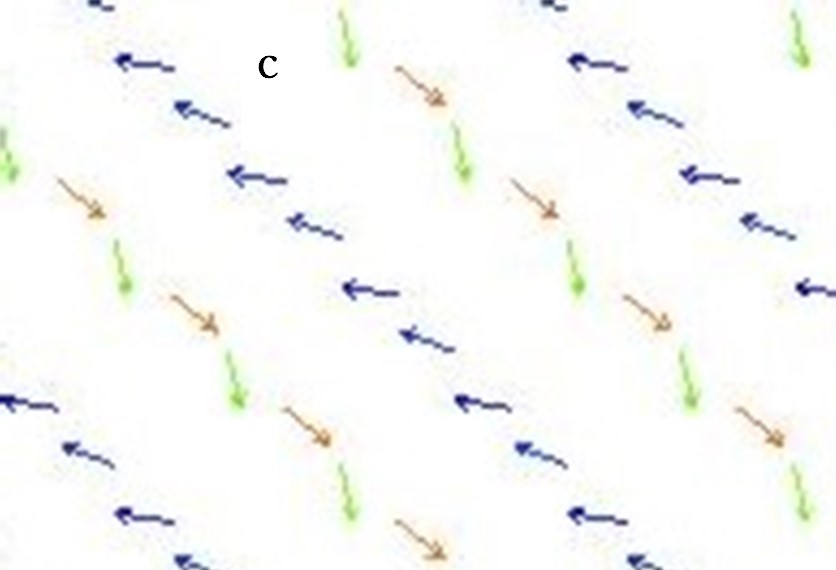}
  \end{center}
 \end{minipage}
 \begin{minipage}{0.24\hsize}
  \begin{center}
   \includegraphics[width=28mm]{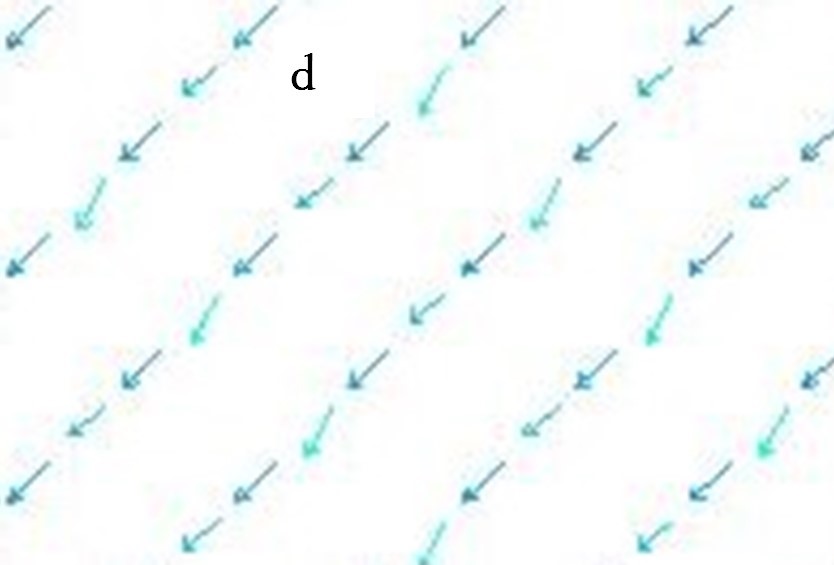}
  \end{center}
 \end{minipage}\vspace{1.5cm}
\vspace{-1.5cm}
\caption{(Color online) Spin orientations in the same region in $xy$ planes of successive $z$ layers for $\omega =0.5$ and $J=0.0$. The color scale is identical to that in Fig. 6.}
\end{figure}
 Spins on sublattices A and D are shown in subfigures b and d, and those on sublattices B and C are shown in subfigures a and c, respectively. The figure shows that spins on sublattice A have period 4 whereas spins on other sublattices have  period 8.

Ordered state for a small exchange interaction $J=0.02$ with $\omega =0.5$ is similar to that for the system with pure dipolar interactions. On the other hand, ordered state for a smaller exchange interaction $J=0.01$ with $\omega =0.5$ is somehow different from that for the pure dipolar system. The peaks of  $\vert S^a ({\bf K})\vert$ are the largest for $a=x$. The largest peak is at ${\bf K}=\pi (0, 0, \delta)$ and the peak height is almost twice as heigh compared to the pure dipolar system. Other peaks of which the peak height is comparable to the pure dipolar system are at ${\bf K}\simeq \pi (1/2, 1/2, 1/2)$, $\pi (1/2, 1/2, -1/2)$, $\pi (1/2, -1/2, 1/2)$, and $(-1/2, 1/2, 1/2)$ within the deviation of $\pi\cdot\delta$. This indicates the coexistence of ferromagnetic order with weaker order of period 4 and two domains related with inversion of magnetization. In Fig. 16, the number distributions of $S^x ({\bf R})$ at sublattices A to D from left to right are depicted.
\begin{figure}
 \begin{minipage}{0.24\hsize}
  \begin{center}
   \includegraphics[width=28mm]{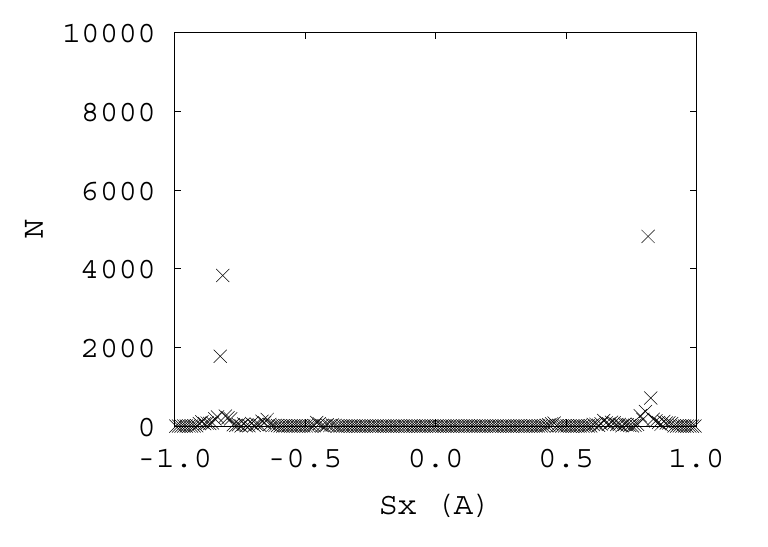}
  \end{center}
 \end{minipage}
 \begin{minipage}{0.24\hsize}
  \begin{center}
   \includegraphics[width=28mm]{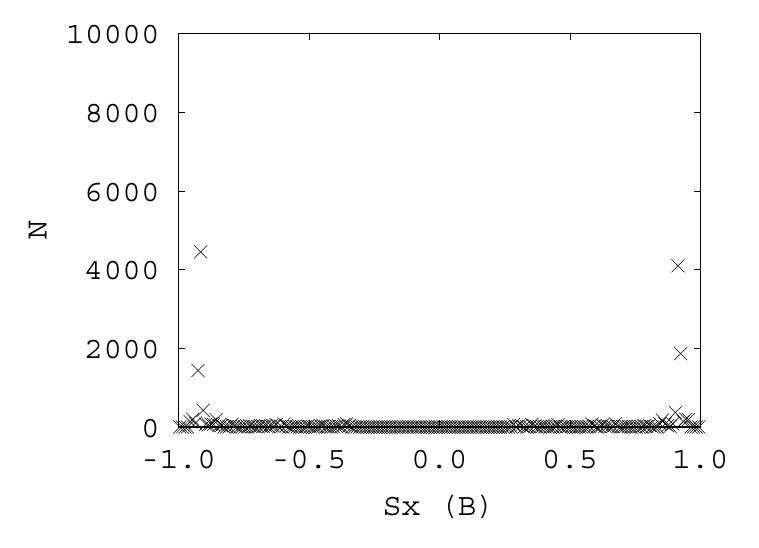}
  \end{center}
 \end{minipage}
 \begin{minipage}{0.24\hsize}
  \begin{center}
   \includegraphics[width=28mm]{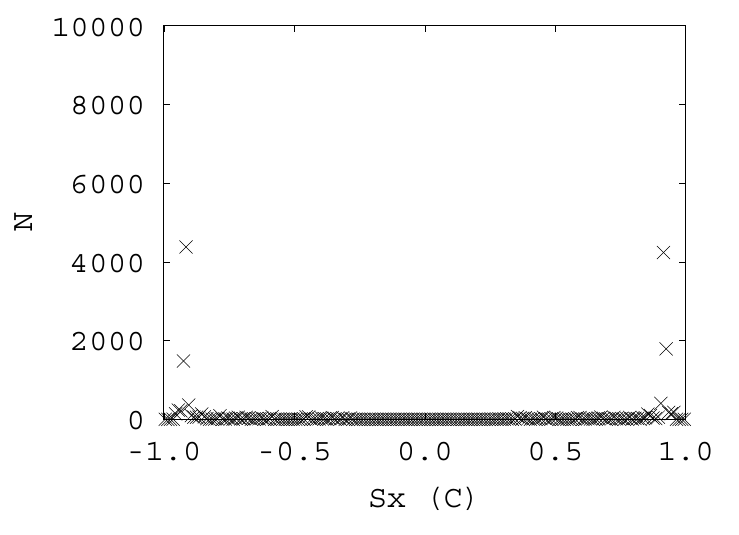}
  \end{center}
 \end{minipage}
 \begin{minipage}{0.24\hsize}
  \begin{center}
   \includegraphics[width=28mm]{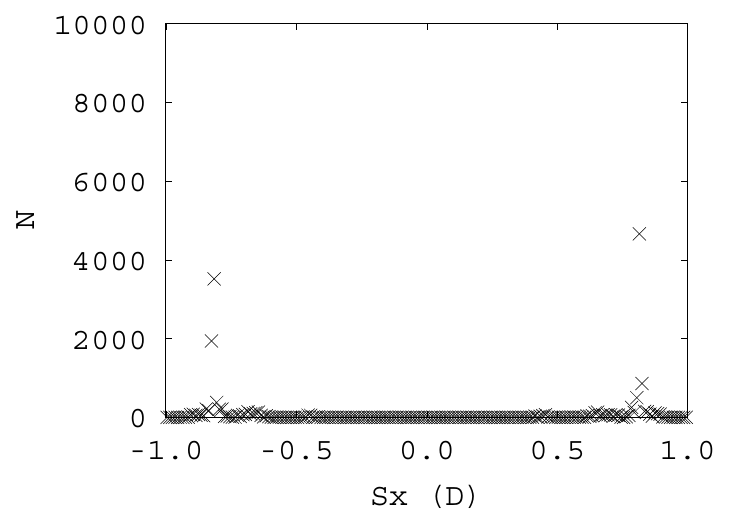}
  \end{center}
 \end{minipage}\vspace{1.5cm}
\vspace{-1.0cm}
\caption{Number distributions of $S^x ({\bf R})$ for $\omega =0.5$ and $J=0.01$ at sublattices A to D from left to right.}
\end{figure}    
Except for the approximate symmetry of the distribution $N(S^x )\simeq N(-S^x )$, there is one peak for each sublattice. This corresponds to the spin structure of period 4. In a magnetic domain, the values for spins belonging to each sublattice are ${\bf S}_{\rm A} \simeq (0.82, 0.42, -0.39)$, ${\bf S}_{\rm B} \simeq (0.92, -0.28, 0.27)$, ${\bf S}_{\rm C} \simeq (0.92, -0.28, -0.27)$, and ${\bf S}_{\rm D} \simeq (0.82, 0.42, 0.39)$. 

\section{Discussion and conclusion}
Heisenberg antiferromagnets on a pyrochlore lattice with long-range dipole-dipole interactions have been investigated by solving the LL equation numerically. Continuous space approximation has been used to take account of long-range nature of the dipole-dipole interactions properly and effectively. Various (ordered) states have been obtained and they are summarized in Table I.
\begin{table}
\caption{Types of order for various $\omega /J$.}
\begin{center}
\begin{tabular}{lll}
\hline
$\omega /J$ & types of order & \\
\hline
0.0 & spin liquid & \\
0.005 & glassy spin state or viscous spin liquid & \\
0.01 & glassy spin state or viscous spin liquid & \\
0.025 & quasi-PC state with small magnetic domains & \\
0.05 & quasi-PC state with small magnetic domains & \\
0.1 & PC state & \\
0.125 & PC state & \\
0.25 & quasi-PC state & \\
0.5 & period 8 & octuple-{\bf K}\\
1.0 & period 8 & octuple-{\bf K}\\
2.5 & period 8 & octuple-{\bf K}\\
5.0 & period 8 & \\
10.0 & period 8 &\\
25.0 & ferromagnetic + period 4 + period 8 (quadruple-{\bf K})& \\
50.0 & ferromagnetic + period 4 (quadruple-{\bf K})& \\
$\infty$ & ferromagnetic + period 4 + period 8 (quadruple-{\bf K})& \\
\hline
\end{tabular}
\end{center}
\end{table}

Without using the continuous space approximation, detailed magnetic structure might be different especially for small range magnetic structures. From the following three examples, the magnetic structures obtained using the approximation seem to be appropriate qualitatively. First, the system with only dipole-dipole interactions has been investigated without using continuous space approximation in Ref. 16. The ordered state is shown to be complex ferromagnetic structure at $T=0.5$ although the complex structure has not been investigated. Appearance of complex ferromagnetic structure is in accord with the result in the present investigation. The validity of the structure with period 4 and (or) 8 appeared in this study should be verified without using the continuous space approximation in the future. Secondly, the PC state, which is a small range magnetic structure, appears for $\omega /J =0.1$ and $0.125$. This parameter range for $\omega /J$ is in accord with that for another investigation\cite{mg2}. Lastly, in the vector spin system on the FCC lattice with only dipole-dipole interactions, the ferromagnetic state is obtained using the continuous space approximation\cite{y3}, which is consistent with the previous study without using the approximation\cite{bz}.

Among the various (ordered) states, spin liquid state\cite{mc} and PC state\cite{pc} have been known for a long time. Here the nature of the spin liquid state is investigated from the spin correlation functions and spin autocorrelation function. PC state seems to appear in a limited range of $\omega /J$\cite{pc}. Quasi-PC states appear for limited ranges of $\omega /J$ with both smaller and larger $\omega /J$.

There are apparently glassy spin state or viscous spin liquid for $0.005\le\omega /J\le 0.01$. As the time scale of dynamical behavior can be very long in these states, it is not clear that the obtained states are precise ground state and should be clarified in the future. Investigating the nature of the glass transition is beyond the scope of the present paper and should also be clarified.

Multiple-{\bf K} states with period 8 have been obtained for rather large values of $\omega /J$ that is $0.5\le\omega /J\le 2.5$. Although multiple-{\bf K} states have been investigated in Ref. 5, the ordered state is different from what have been investigated in this paper. Period 8 states have been obtained for even larger $\omega /J$ that is $\omega /J =5.0$ and $10.0$. For these states, the multiplicity is not clear.

For systems with still larger values of $\omega /J$, ferromagnetic order coexists with periodic order. In the pure dipolar system, magnetic structure with period 4 on one sublattice and period 8 on other three sublattices is realized. Similar magnetic structure appears in the system with $\omega /J =25.0$. In the system with $\omega /J =50.0$, magnetic structure is period 4 on all sublattices. The parameter $\omega /J$ of this system is between the pure dipolar system and the system with $\omega /J =25.0$. At $T=1.0$, the magnetic structure of this system is similar to those for the pure dipolar system and the system with $\omega /J =25.0$. They are similar to the magnetic structure for the pure dipolar system at $T=0.0$ although the peak height of $|S^a ({\bf K})|$ is a little bit lower. For $\omega /J =25.0$, the peaks for the ferromagnetic structure are at ${\bf K}\simeq\pi (0, 0, \pm 3\delta )$ of which the deviation $3\delta$ is three times as large as those for other two systems. The reason why the magnetic structure of the system with $\omega /J = 50.0$ at $T=0.0$ is different from the other two systems is not clear and should be investigated.

\end{document}